# Highly efficient photon-pair source using a Periodically Poled Lithium Niobate waveguide

S. Tanzilli, H. De Riedmatten, W. Tittel, H. Zbinden, P. Baldi,

M. De Micheli, D.B. Ostrowsky, and N. Gisin

*Indexing terms: PPLN, Parametric Down Conversion, Quantum Communication*

We report on a new kind of correlated photon-pair source based on a waveguide integrated on a Periodically Poled Lithium Niobate substrate. Using a pump laser of a few µW at 657 nm, we generate degenerate photon-pairs at 1314 nm. Detecting about 1500 coincidences per second, we can infer a conversion rate of $10^{-6}$ pairs per pump photon, which is four orders of magnitude higher than that obtained with previous bulk sources. These results are very promising for the realization of sources for quantum communication and quantum metrology experiments requiring a high signal-to-noise ratio or working with more than one photon-pair at a time.

*Introduction:* In the beginning of the 80's Alain Aspect [1] performed his famous tests of Bell-inequalities to verify quantum non-locality [2], using a complicated two-photon source based on a double atomic cascade transition. Since then, more and more Bell-tests have been reported [3,4,5], taking advantage of more efficient and handy sources exploiting spontaneous parametric down-conversion (PDC) in second order ($\chi^{(2)}$) non-linear bulk crystals. Such sources have become an essential tool for fundamental quantum optical experiments (like Bell-experiments, teleportation [6,7] or entanglement swapping) as well as for applied fields of research (like quantum key distribution (QKD) [3,8,9] or metrology [10,11]). Although a lot of important results have been obtained, always confirming theoretical predictions, more sophisticated experiments like quantum teleportation suffer from low photon-pair production leading to low signal-to-noise ratios and long measurement times. In this letter, we report on a new kind of twin-photon source taking advantage of an active optical waveguide integrated on a Periodically Poled Lithium Niobate (PPLN) substrate, in opposition to bulk crystals used until now. This leads to an improvement of the pair generation efficiency by four orders of magnitude.

*Characteristics of the PPLN waveguide:* The most efficient parametric generators realized to date consist of Proton Exchange waveguides fabricated in Periodically Poled Lithium Niobate (PPLN) [12], where a periodic reversal of the $\chi^{(2)}$ sign allow Quasi-Phase-Matching (QPM) (see Fig. 1). By an appropriate choice of the inversion period ($\Lambda$), one can quasi-phase-match practically any desired non-linear interaction. Moreover, it becomes possible to use the largest non-linear coefficient of lithium niobate ($d_{33} \approx 30$pm/V), approximately six times larger than the one ($d_{31}$) enabling birefringent phase matching. In addition, the use of a guiding structure permits the confinement of the pump beam over the entire interaction length, rather than simply near the focal point of a lens used for bulk configurations. Our waveguide is made by a new procedure, called Soft Proton Exchange (SPE), with a low-acidity mixture of benzoic acid and lithium benzoate trough a suitable mask. SPE is able to preserve the reversed domains integrity and permits to obtain low propagation losses while leading to a high index variation of about 0.03. Besides increasing the PDC efficiency by several orders of magnitude compared to bulk crystals, a significant advantage for coincidence counting is that the degenerate twin-photons are emitted collinearly in the same mode, making the collection and the separation easier.

*Experimental conditions:* We use a 3.2 cm long sample featuring a 6 µm 1/e width, a 2.1 µm 1/e depth and a poling period of 12.1 µm.

Pumping with a cw laser at 657 nm, QPM conditions for generating degenerate photons at 1314 nm are obtained for a temperature of about 100°C. Heating the sample also allows to cancel the photo-refractive effect in the guide. The bandwidth of the down-converted photons is of about 30 nm Full Width at Half Maximum (FWHM).

The output of the waveguide is coupled to a 50/50 single mode fiber optics beam splitter used to separate the twin photons (see Fig. 1). They are then directed to passively quenched $LN_2$ cooled Germanium Avalanche Photodiodes (Ge-APDs) operated in Geiger mode, showing quantum efficiencies of around 10%. The coincidence rate ($R_C$) is obtained using a Time to Amplitude Converter (TAC) and a Single Channel Analyzer (SCA) [5].

*Measurements and results:* Pumping the guide with 5.2 µW (measured in front of the coupling objective $O_C$), **we obtain an average of 1550 net coincidences per second, and single count rates of 177 kHz including dark count rates of about 22 kHz**. Taking into account 15% coupling of the pump into the guide, **less than 1 µW is used to create the photon pairs**.

Our first figure of merit is the *ratio of coincidence rate ($R_C$) to pump power ($P_P$)*. As shown in table 1, our source features the same order of magnitude for the coincidence rate using $10^3$ times less pump power than other sources reported lately [3,5,13,14].

Second, in order to get a figure of merit independent of the detection efficiencies and the pump power, we can calculate the *conversion efficiency* in terms of photon-pairs created per pump photon. For that, we have to estimate the photon-pair production rate N. The net single count rates $S_1$ and $S_2$ (without dark counts) are then given by:

$$S_1 = \mu_1 \cdot \eta_1 \cdot N \quad (1)$$

$$S_2 = \mu_2 \cdot \eta_2 \cdot N \quad (2)$$

where $\eta_i$ and $\mu_i$ ($i \in \{1,2\}$) are the detection and the collection efficiency for each detector, respectively. Note that $\mu_i$ include all kinds of losses for the twin photons. The net coincidence rate (without accidental coincidences) is:

$$R_c = \mu_1 \cdot \eta_1 \cdot \mu_2 \cdot \eta_2 \cdot N \quad (3)$$

Finally, by inserting the relations (1) and (2) into (3), we obtain:

$$N = \frac{S_1 \cdot S_2}{R_c} \quad (4)$$

In our case as well as in [5,14], N has to be divided by a factor 2 due to the 50% losses of coincidences at the directional coupler. For our new source, we can estimate a pair production rate of about 7.5 MHz corresponding to a conversion efficiency of about $2*10^{-6}$. To our knowledge, this is at least 4 orders of magnitude higher than any other source reported before. Moreover, the PPLN waveguide is so efficient that the electronics used for coincidence counting are rapidly saturated. This is the reason why only very low pump power is used.

Note that our down-converted photons are produced in one single spatial mode. For certain applications it might be interesting to calculate the conversion efficiency normalized with respect to the spectral as well as to the spatial distribution of the fluorescence light. The latter is automatically ensured when coupling the down converted photons into singlemode optical fibers as done in our experiment as well as in [3,5,14].

*Conclusion:* We reported on the first utilization of a quasi phase matched PPLN waveguide for the generation of entangled photon pairs. For only 1 µW at 657 nm, we measured a pair production rate of around 7.5 MHz corresponding to a conversion rate of more than $2*10^{-6}$. Therefore, PPLN waveguides are well suited for experiments where twin-photon production efficiency remains the critical point. One can think of applications of quantum communication over long distances or experiments requiring simultaneous creation of more than one pair, even using conventional pumps such as small and handy picosecond semiconductor laser diodes instead of rather big Ti:sapphire lasers. Finally, it might be possible to build all-pigtailed devices where the PDC source and most of the components needed for the qu-bit treatments are integrated on a single chip. Such systems could open the way to Integrated Quantum Optics.

*Acknowledgement:* We would like to thank the Cost Action P2 "Application of non-linear optical phenomena" and the ESF "Quantum Information Theory and Communications" for financial support. We also thank Douglas Bamford from Gemfire at Palo Alto CA for providing us with the PPLN substrate.

**Author's affiliations:**

S. Tanzilli, P. Baldi, M. De Micheli and D.B. Ostrowsky (Laboratoire de Physique de la Matière Condensée, Université de Nice-Sophia Antipolis, Parc Valrose - 06108 Nice, Cedex 2, France, *tanzilli@unice.fr*)

H. De Riedmatten, W. Tittel, H. Zbinden and N. Gisin (Group of Applied Physics, Université de Genève, 20, rue de l'école de médecine, 1211 Genève 4, Switzerland)

**Figure captions:**

**Fig. 1** *Experimental setup for coincidence counting using a PPLN waveguide.*

$\Lambda$ is the poling period; $\lambda_p$, $\lambda_s$ and $\lambda_i$ are the pump, signal and idler wavelengths; $O_C$ and $O_D$ are the coupling and decoupling objectives; F : pump beam filter; L is the coupling lens into a singlemode fiber; C is a 50/50 directionnal coupler; $D_1$ and $D_2$ are Ge-APDs.

**Table 1.** *Comparison between some of the existing twin-photon sources.*

QPM-PPSF: Quasi Phase-Matched Periodically Poled Silica Fiber

$P_P$ is the pump power. Note that for the PPLN waveguide and the PPSF, the pump power values are guided ones and estimated for the fundamental spatial mode.
$R_C$ is the net coincidence rate; $S_i$ ($i \in \{1,2\}$) are the net single count rates.

***h*** *is the conversion efficiency, i.e. the photon pair production rate per pump photon.* Note that for ref. [13], $\eta$ is given including all the down-converted spatial modes.

Figure 1

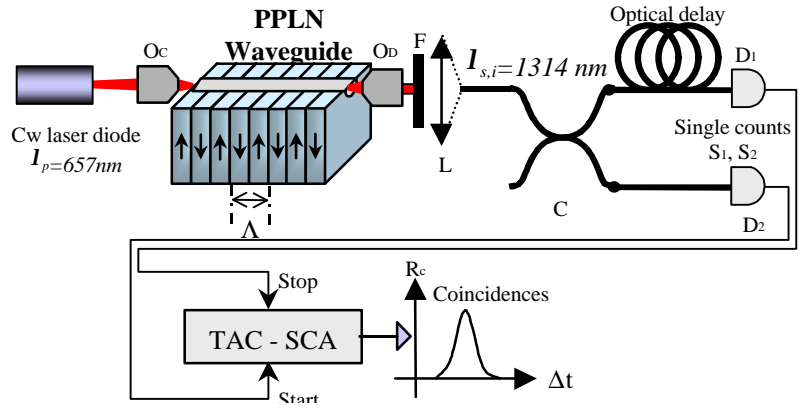

Table 1

|  | PPLN waveguide | KNbO$_3$ bulk [5] | Cascade BBO bulk [13] | Type II-BBO bulk [3] | QPM-PPSF guided [14] |
|---|---|---|---|---|---|
| $P_P$ (mW) | 0.001 | 10 | 150 | 150 | 300 |
| $\lambda_{Pump}$ - $\lambda_{s,i}$ (nm) | 657 - 1314 | 655 - 1310 | 351 - 702 | 351 - 702 | 766 - 1532 |
| APD - det. eff. (%) | Ge - 10 | Ge - 10 | Silicon - 65 | Silicon - 35 | InGaAs - 10 |
| $S_i$ (kHz) | 150 | 250 | 435 | 100 | 36 |
| $R_C$ net (c/s) | 1550 | 5000 | 21000 | 20000 | 500 |
| **$R_C/P_P$** | **$1.6*10^9$** | **$5*10^5$** | **$1.4*10^5$** | **$5.7*10^4$** | **1700** |
| **η** | **$2.2*10^{-6}$** | **$1.9*10^{-10}$** | **$3.4*10^{-11}$** | **$8.1*10^{-13}$** | **$1.1*10^{-12}$** |